# Asymptotic expansions for relativistic celestial mechanics *

M. ARMINJON

*Laboratoire "Sols, Solides, Structures", Institut de Mécanique de Grenoble,
B.P. 53, 38041 Grenoble cedex 9, France. E-mail: arminjon@hmg.inpg.fr*

The method of asymptotic expansions is used to build an approximation scheme relevant to celestial mechanics in relativistic theories of gravitation. A scalar theory is considered, both as a simple example and for its own sake. This theory is summarized, then the relevant boundary problem is seen to be the full initial-value problem. It is shown that, with any given system of gravitating bodies, one may associate a one-parameter family of similar systems, the parameter measuring the gravitational field-strength. After a specific change of units, the derivation of asymptotic expansions becomes straightforward. Two hypotheses could be made as to which time variable has to be used in the expansion. The first one leads to an "asymptotic" post-Newtonian approximation (PNA) with instantaneous propagation, differing from the standard PNA in that, in the asymptotic PNA, all fields are expanded. The second hypothesis could lead to an "asymptotic" post-Minkowskian approximation (PMA) allowing to describe propagation effects, but it is not compatible with the Newtonian limit. It is shown that the standard PNA is not compatible with the application of the usual method of asymptotic expansions as envisaged here.

## 1. Introduction

In order to evaluate the predictions of relativistic theories of gravitation, *e.g.* for the motion of celestial bodies in the solar system, it is necessary to use some approximation method. In Einstein's general relativity (GR), the most widely used method for that purpose is the so-called "post-Newtonian approximation" (PNA), elaborated notably in Refs. [1-5]. The PNA is the method used to consolidate one of the key tests of GR, namely that it predicts so accurately Mercury's residual advance in perihelion. The textbook derivation of that prediction is based on the ideal situation in which Mercury would be a test particle orbiting in the static, spherically symmetric gravitational field of the isolated Sun. Using the PNA, one finds that the same residual advance is predicted if one assumes a more realistic situation where, in particular, the perturbations due to the other planets are accounted for [1, 3, 5]. More generally, the aim of the PNA was to build a consistent relativistic celestial mechanics and to compare its predictions with astronomical observations.

Basically, the standard PNA consists in defining a small parameter $\varepsilon$ that characterizes the weakness of the gravitational field [1, 4], in expanding the *gravitational field* in powers of $\varepsilon$, and in deriving approximate equations from this expansion and the exact field equations. A first question arises: in which sense does the solution of the approximate equations actually approach the relevant solution of the exact equations? This question is complicated, in GR, by the special structure of that theory, in particular by the necessity of imposing a gauge condition [6-7]. To the author's knowledge, this question has not received a clear answer today. A rigorous derivation of a PNA scheme, resulting from the assumed existence of a one-parameter family of metrics and mass tensors obeying the Einstein equations, has yet been proposed by Rendall [8]. But, as noted by Rendall, his scheme differs from the standard PNA in that, in his scheme, all variables, thus gravitational fields as well as matter fields, are expanded.[1] In contrast, in the standard PNA, only the gravitational field is expanded (the leading expansion being that of the space-time metric $\gamma$, which gives the expansions of the Christoffel symbols and the Ricci tensor).[2] As a consequence, Rendall's work [8] does not seem to provide a direct justification for the standard PNA. It rather gives motivation for developing the application of the method of ordinary asymptotic expansion, even if it leads to difficulties in the comparison with observations [12]. Moreover, Rendall [8] does not attempt to describe how the one-parameter family of metrics, that he considers, could be related to a family of physically relevant boundary value problems.

A second difficulty is that the approximate equations for the gravitational field, obtained by using the PNA, turn out to be Poisson equations for the Newtonian potential and some additional, "post-Newtonian" (PN) potentials. Thus, whereas an essential new feature of relativistic theories of gravitation is that gravitation propagates with speed $c$, it is found that gravitation propagates instantaneously at the PNA; this is not only true at the first ($1/c^2$) iteration[3], but also at the second ($1/c^4$) one [13]: only at the "2½" ($1/c^5$) iteration of the PNA does one find out propagation effects, including reaction effects to the emission of gravitational waves [14]. But, as noted by Ehlers *et al.* [6], the determination of the lowest-order time-odd terms in the expansion of the space-time metric, which is crucial in the derivation of the 2½ iteration of the PNA, was not achieved in a self-consistent way in Ref. 14: this determination used results obtained with a different approximation method, namely the standard linearized theory of gravitational radiation; in brief, the statement of the 2½ PNA in Ref. 14 lacks the straightforward character that one might *a priori* expect from an iterative approximation scheme. The "post-Minkowskian" approximation (PMA) schemes differ from the PNA in that the approximate equations for the gravitational field in the PMA involve a d'Alembert (wave) operator in the place of Laplace's operator, thus automatically maintaining the effects of propagation [15-16]. Damour & Schmidt [17] gave existence proofs of one-parameter families of metrics solutions of the Einstein equations, and showed that the successive coefficients in the Taylor expansion (with respect to the parameter) of a such family satisfy the hierarchy of equations encountered in the successive iterations of the PMA. As for Ref. 8 (devoted to the PNA), they do not link the appearance of their family of metrics with a family of physically relevant boundary value problems. Moreover, they study Einstein's *vacuum* field equations, whereas celestial mechanics should consider a system of extended bodies [9]. *In vacuo*, there is only the gravitational field to be expanded, hence the work [17] does not apply directly when the standard PMA, consisting in expanding the gravitational field only [9, 15-16], is used [9] for extended bodies.

The aim of the present paper is to build a general approximation scheme based on the usual method of asymptotic expansions and to obtain the explicit expansions of the fields, equations and boundary conditions up to the order $1/c^2$ included. Preliminary results have been presented in Ref. 18. In this whole work, we shall consider a *scalar,* preferred-frame theory of gravitation, proposed in Refs. 19-20. (*Cf.* Ref. 21 for a review including recent results.) This choice has the following advantages: (i) in the scalar theory, no gauge condition is needed, so that the comparison between the proposed "asymptotic" scheme and the standard PNA will be unbiased; (ii) since the scalar theory assumes a "prior-geometrical" flat metric and since its exact field equation may be formulated in terms of this flat metric, the "flat space-time operators" needed to obtain a tractable approximation scheme occur naturally and do not raise any difficulty of interpretation, in contrast to the situation in GR [6-7]; (iii) the scalar theory is mathematically much simpler than GR, in particular the derivation of asymptotic expansions is very easy. Thus, the choice of this theory provides a simple illustration of the application of the usual method of asymptotic expansion and of its difference with the standard PNA and PMA. Moreover, this theory predicts just the same effects of gravitation on light rays as does GR, thus in the gravitational field of a static spherical body, and this is also true if, as one expects, the body is in a translation with respect to the preferred frame of the theory [22]. Due to the importance of this first experimental check, it is worth to test the scalar theory further and, for that goal, to develop a celestial mechanics according to that theory; the present paper may also be considered as a step in this direction.

---

[1] One must expand all unknown fields if one searches for asymptotic expansions, in the usual sense, of solutions to systems of partial differential equations. Typical examples can be found in transonic gas dynamics [10] and in the homogenization theory as applied, *e.g.,* to porous media or to composite materials [11] .

[2] A *kind* of expansion of some matter variables such as the components of the energy-momentum tensor **T** is used in the standard PNA, but this is a "composite expansion" ; it occurs as a by-product of the expansion of $\gamma$ (due to the fact that any particular expression of **T**, *e.g.* that for a perfect fluid, does involve the metric $\gamma$).

[3] In appropriate units, one may assume that the small parameter $\varepsilon$ is simply $1/c$ with $c$ the velocity of light, as is often formally assumed [see § 5.2].

The plan is as follows: in *Section 2*, we summarize the scalar theory which is considered. In *Section 3*, we derive the exact equations for a perfect fluid. In *Section 4*, we show that the relevant boundary-value problem is the full initial-value problem. In *Section 5*, we describe how to associate, with the given gravitating system S, a one-parameter family ($S_\varepsilon$) of similar systems. In our opinion, this is needed in order to define a meaningful asymptotic approach in a physically relevant situation: if we do not have such family at our disposal, it is rather in a formal sense that we may use an asymptotic method for solving the given physical problem. Of course, one is often led to use formal asymptotics in physics, but the presentation of a framework really suitable for an asymptotic method seems to be a progress. In *Section 6*, we derive the asymptotic expansion of the field equation and the local equations of motion for a perfect fluid, using a PNA and expanding all field variables, and we compare the equations thus obtained, with the equations obtained when one expands the gravitational field only. In *Section 7*, we discuss the boundary conditions for the expanded fields. We finish the paper by the Conclusion (*Section 8*) and a bibliographical note (*Section 9*).

## 2. A short summary of the scalar theory

Space-time is assumed to be equipped with both a flat metric $\gamma^0$ and a curved metric $\gamma$, the latter being more directly related to physical measurements (the "effective metric"). The only equations of this Section that will be used in the remainder are those defining the effective metric in terms of the scalar field $f$ [Eqs.(2.3) and (2.5)], the field equation (2.6) for $f$, and the dynamical equations (2.9)-(2.10). Obviously, they are independent and may be taken axiomatically, independently of the physical status that one wishes to attribute to that theory. The preferred frame E is assumed to be an inertial frame for the flat metric. By this, we mean that there are Galilean coordinates ($x^\mu$) for $\gamma^0$ [*i.e.,* in that coordinates, $(\gamma^0_{\mu\nu}) = (\eta_{\mu\nu}) \equiv \mathrm{diag}(1, -1, -1, -1)$], that are adapted to the frame E. ("Adapted coordinates" are such that any particle bound to the given frame has constant space coordinates [23].) The inertial time $T \equiv x^0/c$ in the preferred frame is called the "absolute time". The scalar field of the theory is the contraction factor $\beta$ affecting the time interval $dt_\mathbf{x}$ measured by a clock fixed at point $\mathbf{x}$ bound to the frame E (due to the dilation of the clock period in the gravitational field), or the square $f \equiv \beta^2$. The latter defines the $\gamma_{00}$ component of the physical space-time metric $\boldsymbol{\gamma}$ in any coordinates $(y^\mu)$, though with $y^0 = cT$, that are adapted to the frame E:

$$dt_{\mathbf{x}}/dT = \sqrt{(\gamma_{00})_E} \equiv \sqrt{f} \equiv \beta. \tag{2.1}$$

The field $f$ is also a potential for the assumed gravity acceleration $\mathbf{g}$ [20]: [4]

$$\mathbf{g} = -\frac{c^2}{2}\nabla f, \quad (\nabla f)^i \equiv g^{0\,ij} f_{,j}, \tag{2.2}$$

where $(g^{0\,ij})$ is the inverse matrix of the component matrix $(g^0_{ij})$ of $\mathbf{g}^0$, with $\mathbf{g}^0$ the spatial part of the flat metric $\boldsymbol{\gamma}^0$ in the frame E (thus $\mathbf{g}^0$ is an invariable Euclidean metric defined on the "preferred reference body" M, *i.e.* the set of the particles bound to E). The field $\beta$ still defines the dilation of an infinitesimal distance $dl$ measured in the direction of vector $\mathbf{g}$, as compared with the distance $dl^0$ evaluated using the Euclidean metric: $dl = dl^0/\beta$, while the infinitesimal distances in directions perpendicular to $\mathbf{g}$ are unaffected. (This makes the physical space metric $\mathbf{g}$ in the frame E discontinuous at the isolated points where $\mathbf{g} = 0$. But $\mathbf{g}$ is merely the effective metric and, moreover, it remains bounded in the neighborhood of any such point: no theoretical difficulty occurs [22].) This gives the following expression for metric $\mathbf{g}$ in the frame E (in any coordinates $(y^\mu)$ adapted to E):

$$g_{ij} = g^0_{ij} + \left(\frac{1}{f}-1\right)h_{ij}, \quad h_{ij} \equiv \frac{f_{,i}f_{,j}}{g^{0\,kl}f_{,k}f_{,l}} \tag{2.3}$$

The space-time metric writes, in any coordinates $(y^\mu)$ adapted to E and such that $y^0 = cT$:

$$ds^2 = \gamma_{\mu\nu}\,dy^\mu\,dy^\nu = f(dy^0)^2 - dl^2, \quad dl^2 = g_{ij}\,dy^i\,dy^j, \tag{2.4}$$

which gives

$$\gamma_{00} = f, \quad \gamma_{ij} = -g_{ij}, \quad \gamma_{0i} = 0. \tag{2.5}$$

Thus, the $\gamma_{0i}$ components are always zero in the ether frame E. This is the expression of the assumption that this frame admits a global simultaneity defined with the absolute time $T$. Except for cosmological problems, the field equation [19] may be written as [20]:

$$\Delta f - \frac{1}{f}\left(\frac{f_{,0}}{f}\right)_{,0} = \frac{8\pi G}{c^2}\sigma \quad (y^0 = cT), \tag{2.6}$$

with $\Delta$ the usual Laplace operator, defined with the Euclidean metric $\mathbf{g}^0$, and where $G$ is Newton's gravitation constant, and $\sigma \equiv (T^{00})_E$ is the mass-energy density in the ether frame. This equation, which is merely space-covariant, implies that gravitation propagates with the velocity of light $c$ (as measured with physical standards affected by gravitation).

Motion is governed by an extension of Newton's second law [20]:

---
[4] Latin indices run from 1 to 3 (spatial indices), Greek indices from 0 to 3.

$$\mathbf{F}_0 + m(v)\mathbf{g} = D\mathbf{P}/Dt_\mathbf{x}. \tag{2.7}$$

With the gravity acceleration **g** assumed (in the preferred frame E) in the theory [Eq.(2.2)], it implies Einstein's motion along geodesics of metric γ, but only for a static gravitational field; however, Newton's second law (2.7) may be defined in any reference frame, and also for metric theories: for those, one adds to the **g** of Eq. (2.2) a certain velocity-dependent term, which gives geodesic motion in the general case [24]. In Eq.(2.7), $\mathbf{F}_0$ is the non-gravitational force, $v$ is the modulus of the velocity **v** of the test particle (relative to the considered arbitrary frame F). The velocity **v** is measured with the local time $t_\mathbf{x}$ synchronized along the given trajectory [25] ($t_\mathbf{x}$ is given by Eq.(2.1) if the preferred frame of the present theory is considered, thus if F = E) and its modulus $v$ is defined with the space metric **h** in the frame F (thus **h** = **g** given by Eq.(2.3) if F = E):

$$v^i \equiv dy^i/dt_\mathbf{x}, \quad v \equiv [\mathbf{h}(\mathbf{v},\mathbf{v})]^{1/2} = (h_{ij}\, v^i v^j)^{1/2}; \tag{2.8}$$

$m(v) \equiv m(0).\gamma_v \equiv m(0).(1 - v^2/c^2)^{-1/2}$ is the relativistic inertial mass, $\mathbf{P} \equiv m(v)\,\mathbf{v}$ is the momentum, and $D/Dt_\mathbf{x}$ is the derivative of a spatial vector appropriate to the case where the Riemannian spatial metric **h** varies with time. In particular, with this derivative, Leibniz' rule for the time derivative of a scalar product $\mathbf{v}.\mathbf{w} = \mathbf{h}(\mathbf{v},\mathbf{w})$ is satisfied [20, 24].

Equations (2.2) and (2.7) define Newton's second law in the preferred frame, for any mass particle. It is thereby defined also for a dust, since dust is a continuum made of coherently moving, non-interacting particles, each of which conserves its rest mass. It then mathematically implies, independently of the assumed form for the space-time metric γ provided it satisfies $\gamma_{0i} = 0$ in the preferred frame, the following dynamical equation for the dust, in terms of its mass tensor $T^{\mu\nu} = \rho^* U^\mu U^\nu$ (with $\rho^*$ the proper rest-mass density and $U^\mu = dy^\mu/ds$ the 4-velocity) [26]:

$$T^{\mu\nu}{}_{;\nu} = b^\mu. \tag{2.9}$$

Here $b_\mu$ is defined by

$$b_0(\mathbf{T}) \equiv \frac{1}{2} g_{jk,0}\, T^{jk}, \quad b_i(\mathbf{T}) \equiv -\frac{1}{2} g_{ik,0}\, T^{0k}. \tag{2.10}$$

[Indices are raised and lowered with metric γ, unless mentioned otherwise. Semicolon means covariant differentiation using the Levi-Civita connection associated with metric γ. The derivation of Eqs.(2.9)-(2.10) [26] is outlined in Ref. 21; they are valid in any coordinates ($y^\mu$) that are adapted to the frame E and such that $y^0 = \phi(T)$ with $T$ the absolute time.] Equation (2.9), with the definition (2.10), is assumed to hold for any material continuum: accounting for the mass-energy equivalence, this is the expression of the universality of gravitation.

### 3. Exact dynamical equations for a perfect fluid

With the help of the identity

$$T^{\mu\nu}{}_{;\nu} = \frac{1}{\sqrt{-\gamma}}\left(\sqrt{-\gamma}\, T^{\mu\nu}\right)_{,\nu} + \Gamma'^{\mu}_{\nu\lambda}\, T^{\nu\lambda} \tag{3.1}$$

[in which the $\Gamma'^{\mu}_{\nu\lambda}$ 's are the Christoffel symbols of metric γ], selecting coordinates such that $\gamma \equiv \det(\gamma_{\mu\nu}) = -1$, and using Eq.(2.5), one may generally rewrite the spatial part of Eq.(2.9) as

$$T^{iv}{}_{,v} + \Gamma^i_{jk} T^{jk} + \frac{1}{2} g^{ij}\left(g_{jk,0} T^{0k} + f_{,j} T^{00}\right) = 0, \qquad (3.2)$$

in which the $\Gamma^i_{jk}$'s are the Christoffel symbols of metric **g** and $(g^{ij}) \equiv (g_{ij})^{-1}$. Note that $\gamma = -1$ in Cartesian coordinates for the Euclidean metric $\mathbf{g}^0$ (*i.e.*, coordinates in which $g^0{}_{ij} = \delta_{ij}$), since Eq.(2.5) implies that $\gamma = fg$ with $g \equiv \det(g_{ij})$ and since the assumed relation between $\mathbf{g}^0$ and the physical space metric **g** implies that

$$g = g^0/f \qquad (g^0 \equiv \det(g^0{}_{ij})). \qquad (3.3)$$

Let us apply Eq.(3.2) to a perfect fluid. The mass tensor is then

$$T^{\mu\nu} = (\mu^* + p/c^2) U^\mu U^\nu - (p/c^2) \gamma^{\mu\nu}, \qquad (3.4)$$

where $\mu^*$ is the volume density of the rest-mass plus internal energy in the proper frame, expressed in mass units,

$$\mu^* \equiv \rho^*(1 + \Pi/c^2). \qquad (3.5)$$

We note that

$$\Gamma^i_{jk} g^{jk} = g^{il} g_{lk,j} g^{jk} - \frac{1}{2} g^{il} g_{jk,l} g^{jk} \qquad (3.6)$$

and that, due to Eq.(3.3), one has

$$g_{jk,l} g^{jk} = \frac{g_{,l}}{g} = \frac{g^0{}_{,l}}{g^0} - \frac{f_{,l}}{f}. \qquad (3.7)$$

Using this, it is straightforward to check that, for a perfect fluid, Eq.(3.2) takes the form

$$\partial_T(\psi u^i) + \partial_j(\psi u^i u^j) + \Gamma^i_{jk} \psi u^j u^k + t^i{}_k \psi u^k = \psi f g^i - g^{ij} p_{,j} \qquad (3.8)$$

(valid in Cartesian coordinates). Here

$$u^i \equiv \frac{dx^i}{dT}, \quad t^i{}_k \equiv \frac{1}{2} g^{ij} \partial_T g_{jk}, \quad \psi \equiv \sigma + p/(c^2 f). \qquad (3.9)$$

As to the "time" part of Eq.(2.9), one gets by lowering index $\mu$ in (2.9) and by writing the $\mu = 0$ component in Cartesian coordinates:

$$\partial_T(\psi f) + \partial_j(\psi f u^j) = (\sigma/2)\partial_T f + (1/c^2)\partial_T p. \qquad (3.10)$$

**4. The question of the relevant boundary value problem for a perfect fluid**

To be definite, we assume henceforth that any matter is in the form of barotropic perfect fluids (one fluid per astronomical body), which means that $\mu*$ in Eq.(3.5), hence also $\rho*$ and $\Pi$, depend only on the pressure $p$, with $\Pi$ being given by [1]

$$\Pi(p) = \int_0^p \frac{dq}{\rho*(q)} - \frac{p}{\rho*(p)}. \tag{4.1}$$

(The integral must be finite.) Note that the source of the gravitational field [Eq.(2.6)] is given by

$$\sigma \equiv (T^{00})_E = \left(\mu* + \frac{p}{c^2}\right)\frac{\gamma_v^2}{f} - \frac{p}{c^2 f} \tag{4.2}$$

and is hence determined by $p$, $f$ and $\mathbf{u}$. The field equation (2.6), together with the dynamical equations (3.8) and (3.10), make a system of five independent partial differential equations for the five independent unknowns $p, f$ and $\mathbf{u}$ (or $\psi, f$ and $\mathbf{u}$; or $\sigma, f$ and $\mathbf{u}$). All of these equations are nonlinear. The dynamical equations (3.8) and (3.10) make a nonlinear first-order system of the same nature as, though more complicated than, the nonlinear system constituted by the continuity equation plus Euler's equation for a perfect fluid in Newtonian gravity (NG) [see Eqs.(5.2) and (5.3) below]. Already in NG, it is not possible to consider that the gravitational field (the potential $U_N$, in NG) is given, i.e. $U_N(\mathbf{x},T)$ known for all $\mathbf{x}$ and $T$; for, in that case, Poisson's equation first determines the density $\rho$, and then Euler's equation alone [though with an initial data $\mathbf{u}(\mathbf{x},T=0)$] determines $\mathbf{u}$, so that the continuity equation has little chance to be satisfied. A similar situation occurs in the new scalar theory, as one easily checks.

However, if the matter source $\sigma(\mathbf{x}, T)$ is considered given, Eq.(2.6) is a quasi-linear second-order equation for the scalar gravitational field $f$. It may be written, in Cartesian coordinates, as

$$A_{\mu\nu} f_{,\mu,\nu} + B = 0, \quad A_{00} \equiv 1/f^2, A_{0i} \equiv A_{i\,0} \equiv 0, A_{ij} \equiv -\delta_{ij}, B \equiv 8\pi G\sigma/c^2 - f_{,0}^2/f^3. \tag{4.3}$$

It is thus a hyperbolic equation, the characteristic equation to be satisfied on a characteristic hypersurface $\phi(x^\mu) = 0$ being [27] $A_{\mu\nu} \phi_{,\mu} \phi_{,\nu} = 0$, i.e. here

$$\frac{1}{c^2 f^2}\left(\frac{\partial \phi}{\partial T}\right)^2 = (\nabla\phi)^2. \tag{4.4}$$

The hypersurface $T = 0$ (or $T$ = Const), i.e. $\phi(x^\mu) \equiv x^0$ − Const = 0, cannot be characteristic, for this would mean that $1/f(\mathbf{x},0)^2 = 0$. Therefore, the Cauchy problem, in which one assumes an arbitrary initial data $f(\mathbf{x},T=0)$ and $\partial_T f(\mathbf{x},T=0)$ for all $\mathbf{x}$, is very plausibly a well-posed problem, and this should be the case for a generic source $\sigma = \sigma(\mathbf{x},T)$. On the other hand, in NG, one also may consider the source $\rho(\mathbf{x}, T)$ as given: but Poisson's equation is an elliptic linear second-order equation for $U_N$, so the problem is well-posed when one assumes, at any time $T$, a spatial boundary condition for $U_N$; typically the condition at infinity, $U_N = O(1/r)$ and $\nabla U_N = O(1/r^2)$ as $r \to \infty$, is assumed.

Now, in NG, the natural boundary value problem for the full system of equations is a mixed initial/ spatial-boundary problem, in which one assumes a spatial boundary condition at any time for $U$, as just mentioned, plus an initial data for the velocity **u** and the density $\rho$ (or the pressure $p$). Since (i) the dynamical equations (3.8) and (3.10) have the same structure as the system made of the continuity equation plus Euler's equation in NG, and (ii) the natural boundary condition for $f$ is an initial condition, whereas the natural boundary condition for $U_N$ is a spatial boundary condition (due to the hyperbolic nature of Eq. (2.6), as opposed to the elliptic nature of Poisson's equation), we are led to consider the full initial-value problem as the relevant boundary value problem in the scalar theory. That is, one should assume the data $\sigma(\mathbf{x},0)$ [or $p(\mathbf{x},0)$, or $\rho*(\mathbf{x},0)$], $\mathbf{u}(\mathbf{x},0)$, $f(\mathbf{x},0)$ and $\partial_T f(\mathbf{x},0)$, for all **x**.

## 5. The situation suitable to derive asymptotic expansions for weak gravitational fields

We consider a gravitating system S made of $N$ separated bodies labeled $a$, $b$ or $c$ (with $a$, $b$, $c$ = 1, 2, ..., $N$), each of which being a spatially compact body made of some material medium, so that the velocity field **v** [Eq.(2.8)] and the field of proper rest-mass density $\rho*$ are unambiguously defined. It is for such a given system, in particular for the solar system, that the celestial mechanics based on some theory of gravitation has to be built and physically tested. Obviously, an agreement between observations and predictions based on arbitrary approximations would prove neither the correctness of the theory nor that of the approximation method. One therefore has to introduce some definite method of approximation, which should have been mathematically tested, if not in the theory of gravitation, at least in other domains. We choose the method of asymptotic expansions, which has been quite extensively tested in many domains (see *e.g.* Kevorkian & Cole [28], Awrejcewicz *et al.* [29], and references therein). Strictly speaking, a consistent asymptotic expansion may be defined and tested only if we have a family $(S_\varepsilon)$ of gravitating systems at our disposal, depending on a parameter $\varepsilon$ which should take values having 0 as an accumulation point. Hence, we should associate a such (virtual) family to the given real system S, with that system corresponding to a given, small value $\varepsilon_0$ of the parameter. Of course, $\varepsilon$ should characterize the weakness of the gravitational field. As is usually done, we shall associate the weak-field limit with a situation in which NG is a good first approximation (indeed an extremely good one, in the solar system). We shall try to introduce this link without logical circularity. As in Section 4, we consider that any matter is in the form of a barotropic perfect fluid. However, we do not need that the state equation, say $\rho* = F(p)$, is the same for each of the $N$ bodies: we merely assume (for simplicity) that $F$ is the same function inside any given body; since the bodies are separated, it is unnecessary to mark this by an index $a$, etc.

### 5.1 *The weak-gravitational-field limit in Newtonian gravity*

In NG, there are five independent fields: $p$, the Newtonian potential $U_N$, and the "flat space-time" velocity, $\mathbf{u} \equiv d\mathbf{x}/dT$. The five independent equations for those fields are: Poisson's equation (of course, there is no distinction in NG between the proper rest-mass density $\rho*$ and the density of rest-mass in the given reference frame, and we shall note both simply $\rho$):

$$\Delta U_N = -4\pi G \rho, \qquad (5.1)$$

the continuity equation:

$$\frac{\partial \rho}{\partial T} + \mathrm{div}(\rho \mathbf{u}) = 0, \qquad (5.2)$$

and Euler's equation,

$$\rho \frac{d\mathbf{u}}{dT} \equiv \rho \left( \frac{\partial \mathbf{u}}{\partial T} + (\text{grad}\,\mathbf{u}).\mathbf{u} \right) = \rho \nabla U_N - \nabla p. \tag{5.3}$$

Let these equations be satisfied for a "Newtonian gravitating system" and, for any $\varepsilon > 0$, set $\xi = \varepsilon/\varepsilon_0$ (where $\varepsilon_0$ will be defined hereafter) and define new fields thus:

$$U_{N\varepsilon}(\mathbf{x}, T) \equiv \xi^2 U_N(\mathbf{x}, \xi T), \quad p_\varepsilon(\mathbf{x}, T) = \xi^4 p(\mathbf{x}, \xi T), \quad \mathbf{u}_\varepsilon(\mathbf{x}, T) = \xi\, \mathbf{u}(\mathbf{x}, \xi T). \tag{5.4}$$

Note that $U_{N\varepsilon_0} = U_N$, $p_{\varepsilon_0} = p$ and $\mathbf{u}_{\varepsilon_0} = \mathbf{u}$. Moreover, in defining

$$F_\varepsilon(p) \equiv \xi^2 F(\xi^{-4} p) \qquad [\text{thus } F_\varepsilon(p) \equiv \varepsilon^2 F_1(\varepsilon^{-4} p)], \tag{5.5}$$

one gets from (5.4)$_2$:

$$\rho_\varepsilon(\mathbf{x}, T) = \xi^2 \rho(\mathbf{x}, \xi T). \tag{5.6}$$

It is immediate to verify that *the fields $p_\varepsilon$, $\mathbf{u}_\varepsilon$, $U_{N\varepsilon}$ and $\rho_\varepsilon$ satisfy also Eqs. (5.1) to (5.3)*. Let us introduce an adimensional field-strength parameter as Misner *et al.* [4]

$$\varepsilon_0 \equiv (U_{N\text{max}}/c^2)^{1/2}, \qquad U_{N\text{max}} \equiv \max\{U_N(\mathbf{x}, T); (\mathbf{x}, T) \in \mathbf{R}^4\}. \tag{5.7}$$

($U_{N\text{max}}$ is assumed finite; this will be the case, *e.g.*, for an "insular mass distribution" quasi-perodic in time; if necessary, one may restrict the time domain to a compact interval.) From (5.4)$_1$, it follows that $[(U_{N\varepsilon})_{\text{max}}/c^2]^{1/2} = \varepsilon$. Thus, with any Newtonian system, we may associate a one-parameter family of systems. All systems $S_\varepsilon$ have just the same geometry, only the time scale increases as $\varepsilon^{-1}$ (accordingly, the velocity decreases as $\varepsilon$), the density and the potential decrease as $\varepsilon^2$, and the pressure as $\varepsilon^4$, as $\varepsilon$ tends towards zero. This is the weak-gravitational-field limit in NG. (Yet the family extends to strong Newtonian fields, for $\varepsilon$ is arbitrary.) This result explains order-of-magnitude estimates partially justified by virial theorems (see *e.g.* Will [5], p. 89) and allows to use them in a definite asymptotic framework.

### 5.2 *A convenient choice of units*

In expanding the field equations (*e.g.*, of the scalar theory), it will be essential to know the order of the various fields. It turns out that, in adopting a specific variation of the units with $\varepsilon$, the situation becomes very simple. The change of units affects merely the time unit, which is multiplied by $\varepsilon^{-1}$ for the system $S_\varepsilon$, and the mass unit, which is multiplied by $\varepsilon^2$. Note that $G$ is not affected by this change. But the time $T$ is changed to $T' \equiv \varepsilon T$, hence the velocity to $\mathbf{u}' \equiv \varepsilon^{-1} \mathbf{u}$ and, in particular, the velocity of light becomes $c' \equiv \varepsilon^{-1} c$ so that, *in the new units, the small parameter may be taken to be $\varepsilon' = 1/c'$* (with $\varepsilon' = \varepsilon$ if $c = 1$ in the starting units). The mass density is changed to $\rho' \equiv \varepsilon^{-2} \rho$, and the Newtonian potential becomes $U_N' \equiv \varepsilon^{-2} U_N$. Finally, the pressure is changed to $p' \equiv \varepsilon^{-4} p$. Therefore, we have in the new units:

$$U_N'{}_\varepsilon(\mathbf{x}, T') = \varepsilon^{-2} U_{N\varepsilon}(\mathbf{x}, T) = \varepsilon^{-2} \xi^2 U_N(\mathbf{x}, \xi T) = \varepsilon_0^{-2} U_N(\mathbf{x}, \varepsilon_0^{-1} T') = U_{N1}(\mathbf{x}, T'), \tag{5.8}$$

and in the same way $\mathbf{u}'_\varepsilon(\mathbf{x}, T') = \mathbf{u}_1(\mathbf{x}, T')$, etc. Thus, in the new units, all fields are independent of $\varepsilon$, and they have precisely the values they have (in the starting units) in the particular system $S_1$, characterized by $\varepsilon = 1$. The spatial, as well as the temporal derivatives, will hence also stay independent of $\varepsilon$ in the new units, hence in particular

$$\frac{\partial U_{N'\varepsilon}}{\partial T'} = \frac{\partial U_{N1}}{\partial T'} = \mathrm{ord}(\varepsilon^0). \tag{5.9}$$

[As noted by Hinch [30], the usual notation $\varphi = O(\varepsilon^k)$ implies only that $\varphi$ is of order lower than or equal to $k$, thus we adopt his notation $\varphi = \mathrm{ord}(\varepsilon^k)$ when $\varphi$ is of order $k$ exactly.] Note, however, that this situation in which the fields $p_\varepsilon$, $\mathbf{u}_\varepsilon$, $U_{N\,\varepsilon}$ and $\rho_\varepsilon$ are just the same (up to a change of units) independently of $\varepsilon$, is special to NG.

### 5.3 Definition of the weak-gravitational-field limit in the scalar theory

As was said at the beginning of this Section, the proposed definition for a weak-gravitational-field limit will consist in defining a family $(S_\varepsilon)$ of systems, from the data of the given gravitational system of interest. Since, as was seen in Section 4, the natural boundary-value problem in the scalar theory is the initial-value problem, we have to define a family of initial conditions. Let us first state the two conditions that should be fulfilled by the family $(S_\varepsilon)$:

(i) The curved metric $\boldsymbol{\gamma}$ should tend towards the flat metric $\boldsymbol{\gamma}^0$ as $\varepsilon \to 0$.

(ii) The fields $\rho*$, $\Pi$, $p$ and $\mathbf{u}$ should be of the same order in $\varepsilon$ as in the weak-field limit of NG, hence should be $\mathrm{ord}(\varepsilon^0)$ in the varying units of §5.2.

In view of Eqs.(2.3)-(2.4), condition (i) amounts to say that the dimensionless scalar field $f = f_\varepsilon$ tends towards 1 as $\varepsilon \to 0$. Therefore, we are led to assume an expansion

$$f_\varepsilon = 1 + \phi\,\varepsilon^k + O(\varepsilon^{k+1}). \tag{5.10}$$

We must determine the integer $k$. To this end, we shall enter this expansion into the field equation (2.6). We first get for the spatial part:

$$\Delta f_\varepsilon = \varepsilon^k \Delta \phi + O(\varepsilon^{k+1}),$$

because the differentiation with respect to the spatial variables does commute with the expansion with respect to $\varepsilon$, under certain uniformity assumptions for the expansions [ensuring that the spatial derivatives of the remainder term in (5.10) are still $O(\varepsilon^{k+1})$]. The two foregoing equations are valid in the fixed units as well as in the varying units, because the length unit is unchanged. To achieve the determination of $k$, we adopt the varying units, such that the small parameter $\varepsilon$ is proportional to $\varepsilon' = 1/c'$. But therefore, *we have to precise whether the time variable is $T' = \varepsilon T$, or $c'\,T' = \varepsilon T/\varepsilon' \propto T$. This is exactly the departure point between the PNA (time variable $T'$) and the PMA (time variable $c'\,T'$)*, as we shall see. Accordingly, in a PNA (resp. in a PMA), we must assume that the derivative, with respect to $T'$ (resp. with respect to $c'T'$), of remainder terms like $O(\varepsilon^{k+1})$ in (5.10), is still $O(\varepsilon^{k+1})$. Under this assumption, we find that the term involving time derivatives in Eq.(2.6) is $O(\varepsilon^{k+2})$ [resp. $O(\varepsilon^k)$]. Hence the *left-hand side* of (2.6) is $\mathrm{ord}(\varepsilon^k)$. On the right-hand side, $\sigma$ is given by Eq.(4.2). Hence, we get $\sigma_\varepsilon = \mathrm{ord}(\varepsilon^0)$ by condition (ii), therefore the *right-hand side* of (2.6) is $\mathrm{ord}(\varepsilon^2)$. Thus *one must have $k = 2$ in Eq.(5.10)*. We may set (in any units):

$$f_\varepsilon = 1 - 2V_\varepsilon/c^2, \qquad V_\varepsilon/c^2 = \mathrm{ord}(\varepsilon^2) \tag{5.11}$$

(the coefficient $-2$ is there for convenience; the ratio by $c^2$ makes $V$ homogeneous to the Newtonian potential). Eq. $(5.11)_1$ defines $V_\varepsilon$, the important point is the estimate $(5.11)_2$.

Next, we recall a recent argument [18] showing that *the time variable in the varying units must be T'*, *i.e.,* a PNA must be chosen. The first point is that the asymptotic expansions will have first of all to be valid at *fixed* values of the chosen time and space variables. Since we choose to formulate the expansions in the varying units, it is then not equivalent to assume expansions with $x'^0 = T'$ as the time variable (PN expansions) or with $x^0 = c'T'$ (PM expansions), because the ratio $x'^0/x^0 = 1/c' = \varepsilon/c$ is proportional to the small parameter: it is easy to convince oneself that a PN expansion can not generally be rewritten as a PM one, and *vice-versa* [18]. Moreover, in the varying units, the time variable is $T' = \varepsilon T$ where $T$ is the "true" time (*i.e.* that measured in fixed units), hence $c'T' = cT$ is (proportional to) the true time. But since the true orbital velocities in the system $S_\varepsilon$ vary like $\varepsilon$, the true orbital periods are like $\varepsilon^{-1}$. Hence it is $T'$, not $c'T' \propto T$, which remains nearly the same, as $\varepsilon$ is varied, for one orbital period of a given body (*a*). In the PMA, the time variable would be the true time $T = T'/\varepsilon$ and one thus should have at fixed **x** and $T$, if $\rho_\varepsilon$ is ord($\varepsilon^0$) in the varying units [as imposed by condition (ii)]:

$$\rho'_\varepsilon(\mathbf{x}, T) = \rho_0(\mathbf{x}, T) + O(\varepsilon)$$

(where index 0 means zero-order coefficient in the expansion). But this means that the position of the bodies is approximately independent of $\varepsilon$ at a fixed true time, in contradiction with the assumption that the orbital velocities vary like $\varepsilon$. Finally, the reason why the choice of $T'$ (resp. of $c'T'$) as the time variable in the expansions may be called a PNA (resp. a PMA) is that the first choice leads to Poisson equations for the gravitational potentials, whereas the second one leads to wave equations [see Eqs. (6.6)-(6.8) below]. However, we have just seen that the second choice is not allowed in the "near zone" occupied by the bodies.

Now we define the family of systems, *i.e.* the fields $f_\varepsilon$, $p_\varepsilon$, $\mathbf{u}_\varepsilon$, and $\rho^*_\varepsilon$, from the data of one gravitating system of perfect-fluid bodies in the scalar theory: we assume that the fields $f$, $p$, $\mathbf{u}$, and $\rho^* = F(p)$, are known (for all $\mathbf{x} \in M$ and in some interval $I = [0, T_1]$ for $T$). We provisionally come back to the invariable units for clarity, and we set [*cf.* Eq. (5.11)]

$$V \equiv c^2(1-f)/2, \tag{5.12}$$

$$V_{max} \equiv \max\{V(\mathbf{x}, 0); \mathbf{x} \in M\}, \quad \varepsilon_0 \equiv (V_{max}/c^2)^{1/2}, \tag{5.13}$$

$\xi = \varepsilon/\varepsilon_0$. We define $f_\varepsilon$, $p_\varepsilon$, $\mathbf{u}_\varepsilon$ [and $\rho^*_\varepsilon = F_\varepsilon(p_\varepsilon)$ with $F_\varepsilon$ given by (5.5)], as the solution (assumed unique for all $\mathbf{x} \in M$ and in some interval $[0, T_2(\varepsilon)]$ for $T$) of the initial value problem

$$f_\varepsilon(\mathbf{x}, 0) = 1 - 2\xi^2 \frac{V(\mathbf{x}, 0)}{c^2}, \quad \frac{\partial f_\varepsilon}{\partial T}(\mathbf{x}, 0) = -\frac{2\xi^3}{c^2} \frac{\partial V}{\partial T}(\mathbf{x}, 0), \tag{5.14}$$

$$p_\varepsilon(\mathbf{x}, 0) = \xi^4 p(\mathbf{x}, 0), \qquad \mathbf{u}_\varepsilon(\mathbf{x}, 0) = \xi \mathbf{u}(\mathbf{x}, 0), \tag{5.15}$$

for the field equations (2.6), (3.8) and (3.10). Equation (5.14)$_1$ means that, in Eq.(5.11)$_1$,

$$V_\varepsilon(\mathbf{x}, 0) = \xi^2 V(\mathbf{x}, 0). \tag{5.16}$$

This is the simplest way to get the estimate (5.11)$_2$. Also, Eqs. (5.15) are the simplest way to ensure that the fields $p_\varepsilon$ and $\mathbf{u}_\varepsilon$ are ord($\varepsilon^4$) and ord($\varepsilon$) respectively, as in the weak-field limit of NG. Moreover, $\mathbf{u}_\varepsilon = \mathrm{ord}(\varepsilon)$ means an $\varepsilon^{-1}$ dependence of the characteristic time; if one combines this with Eq.(5.11), one expects the asymptotic validity of Eq.(5.4)$_1$ for $V$:

$$V_\varepsilon(\mathbf{x}, T) \sim \xi^2 V(\mathbf{x}, \xi T) \text{ as } \varepsilon \to 0. \qquad (5.17)$$

In uniform conditions, one may differentiate (5.17) with respect to $T$. This leads to define $\partial_T f_\varepsilon(\mathbf{x}, 0)$ by Eq.(5.14)$_2$. Thus, the initial data (5.14)-(5.15) is justified by (5.11), which follows from conditions (i) and (ii), and by the wish that the fields $p_\varepsilon$ and $\mathbf{u}_\varepsilon$ behave as in the weak-field limit of NG, *i.e.* by condition (ii). In other words, the minimal conditions (i) and (ii) imposed to the weak-field limit lead essentially uniquely to the initial data (5.14)-(5.15). Conversely, one expects that the initial data (5.14)-(5.15) ensures that equation (5.17) holds and that

$$p_\varepsilon(\mathbf{x}, T) \sim \xi^4 p(\mathbf{x}, \xi T) \text{ and } \mathbf{u}_\varepsilon(\mathbf{x}, T) \sim \xi \mathbf{u}(\mathbf{x}, \xi T) \text{ as } \varepsilon \to 0, \qquad (5.18)$$

whence also, the state equation being $\rho^*_\varepsilon = F_\varepsilon(p_\varepsilon)$ with $F_\varepsilon$ given by (5.5),

$$\rho^*_\varepsilon(\mathbf{x}, T) \sim \xi^2 \rho^*(\mathbf{x}, \xi T). \qquad (5.19)$$

Thus, we have associated with the physically given system S a one-parameter family of similar systems, also governed by the equations of the scalar theory. This provides a framework in which one might test the convergence of the asymptotic expansions derived hereafter. Once again, the essential point is that, in deriving asymptotic expansions, the existence of a one-parameter family (S$_\varepsilon$) is implicitly assumed: it seems better to have at least a (very) good candidate for such family. Moreover, the initial data (5.14)-(5.15) will allow us to state initial and boundary conditions for the expanded fields (Section 7). Note that Eqs.(5.10) to (5.19) hold true in the varying units, if one also expresses in these units the velocity of light and the fields $V$, $p$, $\mathbf{u}$ and $\rho^*$, corresponding to the physically given gravitating system (which makes those fields *depend* on $\varepsilon$ ...).

Since Eqs.(5.17)-(5.19) may be assumed to be valid, all fields, *e.g.* $p_\varepsilon$, $\mathbf{u}_\varepsilon$, $V_\varepsilon$ and $\rho^*_\varepsilon$, are ord($\varepsilon^0$) in the specific varying units, in which $\varepsilon \propto 1/c$. This means that our condition (ii), as well as [entering (5.17) into (5.12)] our condition (i) are indeed verified. It also makes the derivation of asymptotic expansions a routine affair. There is no theoretical difficulty in considering expansions with an arbitrary number $n$ of terms but, of course, the explicit expressions become very involved as $n$ increases. Since, in the solar system, the second iteration, *i.e.*, that immediately following the Newtonian approximation, leads merely to minute corrections, we shall stop at this second iteration. As a last preliminary point before expanding the fields in powers of $\varepsilon$, one has to ask a question about the parity of the expansions. We proved that the first two terms in the expansion of the scalar field are even powers of $\varepsilon \propto 1/c$ [Eq.(5.11)]. But a look at the basic equations for a perfect fluid in the scalar theory, Eqs.(2.6), (3.8) and (3.10), shows that *it is only the square $1/c^2$ that plays a role (thus, the actual small parameter is $1/c^2$)*, hence more generally *one may assume even powers in $1/c$*. In Ref. 18, some test of this assumption has been performed: it has been checked that the first-order expansion in $1/c$ gives the same results as the zero-order expansion in $1/c^2$ - namely, both expansions are equivalent to Newtonian gravity (NG). However, an important comment is in order: polynomials in the effective small parameter $(1/c^2)$, *i.e.* Taylor expansions, are merely the most obvious possibility for asymptotic expansions, corresponding to the most regular behaviour. It may well be that these simple expansions work only until some finite order. Thus, in GR, there is a breakdown in the approximations based on even-powers expansions, either at the order $1/c^4$ or at the order $1/c^6$, depending on the gauge condition [8]. May be one should assume *then* new expansion functions, *e.g.* odd powers in $1/c$ - as is done for describing the first radiation effects in the standard PNA, where an $1/c^5$ power is introduced [14].

### 6. Expansion of all fields vs. expansion of the gravitational field only

In this Section, we shall assume that, in the varying units, the fields admit expansions with $T$ (formerly $T'$ : *we omit the prime henceforth*) as the time variable; this may be called the "PN hypothesis". We begin with an ordinary asymptotic expansion, then we show that the standard PNA is a different kind of expansion.

    6.1 *Ordinary asymptotic expansion: all fields expanded*
    Since we assume only even powers in $1/c$, the independent fields are expanded as follows:

$$f = 1 - 2\,U/c^2 - 2\,A/c^4 + O(\varepsilon^6), \tag{6.1}$$

$$p = p_0 + p_1/c^2 + O(\varepsilon^4), \tag{6.2}$$

$$\mathbf{u} = \mathbf{u}_0 + \mathbf{u}_1/c^2 + O(\varepsilon^4). \tag{6.3}$$

(*Henceforth, we omit the index $\varepsilon$ for $f_\varepsilon$, etc.*) These equations are written in the (specific) varying units. The coefficients of the expansions are the fields $U$, $A$, $p_0$, $p_1$, $\mathbf{u}_0$, $\mathbf{u}_1$. By definition of an asymptotic expansion, these fields are independent of $\varepsilon$ or $1/c$ (in those varying units), and they are also independent of the order of the expansion: in the next approximation, we add a term to any expansion, *e.g.* we add $p_2/c^4$ to Eq.(6.2), but the fields of the preceding approximations, *i.e.* $p_0$ and $p_1$, are left unchanged. In invariable units, Eq.(6.1) holds true [although $U$ becomes ord($\varepsilon^2$) and $A$ becomes ord($\varepsilon^4$)], whereas Eqs.(6.2) and (6.3) change to

$$p = \varepsilon^4\,[p_0 + \varepsilon^2\,p_1/c^2 + O(\varepsilon^4)], \qquad \mathbf{u} = \varepsilon\,[\mathbf{u}_0 + \varepsilon^2\,\mathbf{u}_1/c^2 + O(\varepsilon^4)], \tag{6.4}$$

where $p_0$, $p_1$, etc., have the same numerical values as in the varying units (for the corresponding values of the respective time variables), hence do not depend on $\varepsilon$. Having expanded the independent fields, we get by Taylor's formula an expansion of the other fields, *e.g.*

$$\rho^* = \rho^*_0 + \rho^*_1/c^2 + O(\varepsilon^4), \quad \rho^*_0 \equiv F_1(p_0), \quad \rho^*_1 \equiv F_1'(p_0).p_1 \tag{6.5}$$

[the state equation has been assumed to be, in invariable units, $\rho^*_\varepsilon = F_\varepsilon(p_\varepsilon)$ with $F_\varepsilon$ given by (5.5); in the varying units, this becomes $\rho^*_\varepsilon = F_1(p_\varepsilon)$].

    Let $x^0$ be the time variable used in the expansions. (Thus $x^0 = T$ actually, and $x^0 = cT$ for a PM expansion, even though we know from §5.3 that the latter is incompatible with the assumption that the true orbital velocities vary like $\varepsilon$.) We get from (6.1):

$$\Delta f = \frac{-2}{c^2}\Delta U - \frac{2}{c^4}\Delta A + O(\varepsilon^6), \tag{6.6}$$

$$-\frac{1}{f}\left(\frac{f_{,0}}{f}\right)_{,0} = \frac{2}{c^2}U_{,0,0} + \frac{2}{c^4}\left(A_{,0,0} + 2U_{,0}^2 + 4UU_{,0,0}\right) + O(\varepsilon^6). \quad (6.7)$$

Since $x^0 = T$, the field equation (2.6) gives with (6.6) and (6.7):

$$\frac{2}{c^2}\left[-\Delta U + \frac{1}{c^2}\left(-\Delta A + \frac{\partial^2 U}{\partial T^2}\right) + O(\varepsilon^4)\right] = \frac{2}{c^2}\left[4\pi G\left(\sigma_0 + \frac{\sigma_1}{c^2} + O(\varepsilon^4)\right)\right], \quad (6.8)$$

where on the r.h.s. we have replaced $\sigma$ [Eq.(4.2)] by its asymptotic expansion, which will be deduced a few lines later from the expansions (6.1)-(6.3). Thus, $\sigma_0$ and $\sigma_1$ do *not* depend on $\varepsilon$, hence we may identify the powers to get the *separate* and *exact* equations

$$\Delta U = -4\pi G \sigma_0, \quad (6.9)$$

$$\Delta A = -4\pi G \sigma_1 + \partial^2 U/\partial T^2. \quad (6.10)$$

These two Poisson equations imply an instantaneous propagation of gravitation, as it is characterized at the order $n = 2$ by the fields $U$ and $A$. It is easy to convince oneself that the same occurs at *any* approximation, *i.e.* at any order $n$ of the expansion, under the "PN hypothesis" $x^0 = T$. It is also easy to verify that the "PM hypothesis" $x^0 = cT$ would lead to wave equations with propagation at speed $c$ – but the PM hypothesis is untenable in the near zone (§5.3 and Ref. 18).

The spatial metric (2.3) is expanded, using (6.1), to

$$g_{ij} = \delta_{ij} + \frac{2U}{c^2}h^1_{ij} + O(\varepsilon^4), \quad g^{ij} = \delta_{ij} - \frac{2U}{c^2}h^1_{ij} + O(\varepsilon^4), \quad (6.11)$$

where

$$h^1_{ij} \equiv \frac{U_{,i}U_{,j}}{U_{,k}U_{,k}}. \quad (6.12)$$

We introduce the density of rest mass in the preferred frame and with respect to the Euclidean volume measure $dV \equiv \sqrt{g^0}\, dy^1\, dy^2\, dy^3$. This density (denoted by $\rho_{00}$ in previous works, while $dV$ was denoted by $dV^0$ instead [20-21, 26]) is defined by [20]:

$$\rho \equiv dm_0/dV \equiv \rho^* \gamma_v/\beta. \quad (6.13)$$

Its expansion is given by $\rho = \rho_0 + \rho_1/c^2 + O(\varepsilon^4)$, with

$$\rho_0 = \rho^*_0, \qquad \rho_1 = \rho^*_1 + \rho_0(\mathbf{u}_0^2/2 + U). \quad (6.13\text{bis})$$

Using this expansion we get by entering the expansions (6.2), (6.3) and (6.11)$_1$ into Eqs. (4.2) and (3.9):

$$\psi_0 = \sigma_0 = \rho_0 = \rho^*_0, \quad \sigma_1 = \rho_1 + \rho_0(\mathbf{u}_0^2/2 + \Pi_0 + U), \quad \psi_1 = \sigma_1 + p_0. \quad (6.14)$$

The expansion of the energy equation (3.10) is then found to be

$$\partial_T \rho_0 + \partial_j(\rho_0 u_0{}^j) = 0, \qquad (6.15)$$

$$\partial_T\left[\rho_0\left(\frac{\mathbf{u}_0{}^2}{2} + \Pi_0 - U\right) + p_0 + p_1\right] + \partial_j\left\{\left[\rho_0\left(\frac{\mathbf{u}_0{}^2}{2} + \Pi_0 - U\right) + p_0 + p_1\right]u_0{}^j + \rho_0 u_1{}^j\right\} =$$
$$= -\rho_0 \, \partial_T U + \partial_T p_0. \qquad (6.16)$$

The local equations of motion (3.8) are expanded to

$$\partial_T(\rho_0 u_0{}^i) + \partial_j(\rho_0 u_0{}^i u_0{}^j) = \rho_0 U_{,i} - p_{0,i}, \qquad (6.17)$$

$$\partial_T(\rho_0 u_1{}^i + \psi_1 u_0{}^i) + \partial_j(\psi_1 u_0{}^i u_0{}^j + \rho_0 u_1{}^i u_0{}^j + \rho_0 u_0{}^i u_1{}^j) + {}_1\Gamma^i_{jk}\rho_0 u_0{}^j u_0{}^k + \rho_0 u_0{}^j \partial_T k_{ij} =$$
$$= -p_{1,i} + 2 k_{ij} p_{0,j} + \psi_1 U_{,i} + \rho_0 A_{,i} - 2U\rho_0 U_{,i}, \qquad (6.18)$$

where

$${}_1\Gamma^i_{jk} \equiv k_{ij,k} + k_{ik,j} - k_{jk,i}, \qquad k_{ij} \equiv U h^1{}_{ij}. \qquad (6.19)$$

Now Eqs.(6.15) and (6.17) are just the Newtonian equations applied to the zero-order expanded fields of the scalar theory: Eq.(6.15) is the continuity equation (5.2); with Eq.(6.17), it implies Euler's equation (5.3), thus

$$\partial_T(u_0{}^i) + u_0{}^i{}_{,j} u_0{}^j = U_{,i} - \frac{p_{0,i}}{\rho_0}. \qquad (6.20)$$

They imply the validity of the Newtonian energy equation:

$$\partial_T\left[\rho_0\left(\frac{\mathbf{u}_0{}^2}{2} + \Pi_0 - U\right)\right] + \partial_j\left\{\left[\rho_0\left(\frac{\mathbf{u}_0{}^2}{2} + \Pi_0 - U\right) + p_0\right]u_0{}^j\right\} + \rho_0 \partial_T U = 0. \qquad (6.21)$$

Using the latter, the second-order energy equation, Eq.(6.16), simplifies to

$$\partial_T \rho_1 + \partial_j(\rho_1 u_0{}^j + \rho_0 u_1{}^j) = 0. \qquad (6.22)$$

The latter shows that the rest-mass is conserved at the PNA of the scalar theory: together with Eq.(6.15), Eq.(6.22) implies that

$$\frac{\partial \rho}{\partial T} + \text{div}(\rho \mathbf{u}) = O(\varepsilon^4). \qquad (6.23)$$

But the rest-mass is not exactly conserved for a perfect fluid. Indeed a *creation/destruction rate* is predicted by the scalar theory, in a variable gravitational field under a fluid pressure. For a weak field, this rate is ord($\varepsilon^4$) and compatible with the experimental evidence on mass conservation [26]. We also note that the equations for the first PN corrections, *i.e.* the equations for the order $1/c^2$, Eqs.(6.10), (6.18) and (6.22), *are all linear with respect to the "PN fields"* $A$, $p_1$, $\psi_1$, $\mathbf{u}_1$. This is typical of a perturbation method.

### 6.2 *Standard PNA: gravitational field alone expanded*

If we apply to the scalar theory the standard PNA, as exposed by Fock [1], Chandrasekhar [2], Weinberg [3], Misner *et al.* [4], or Will [5], then we expand the scalar field $f$, Eq.(6.1), but we leave $p$ and $\mathbf{u}$ "unexpanded". Then Eqs.(6.6) and (6.7) still allow to write the l.h.s. of the field equation (2.6) as the l.h.s. of Eq.(6.8). If we wish, we even may express the matter source $\sigma$, Eq.(4.2), as a *kind* of expansion:

$$\sigma = \sigma_0 + \sigma_1/c^2 + O(\varepsilon^4), \qquad \sigma_0 \equiv \rho, \qquad \sigma_1 \equiv \rho\left(\mathbf{u}^2/2 + \Pi + U\right), \tag{6.24}$$

so that the whole of Eq.(6.8) holds true with this new definition of $\sigma_0$ and $\sigma_1$. But since $\rho$, $\mathbf{u}$ and $\Pi$ are unexpanded and hence (of course) depend on $\varepsilon$, the coefficients of this "expansion", $\sigma_0$ and $\sigma_1$, now *do depend* on $\varepsilon$ (this is called a "composite expansion"). As a consequence, we may only rewrite Eq.(6.8) as the *approximate* and *unseparate* equation

$$-\Delta U + \frac{1}{c^2}\left(-\Delta A + \frac{\partial^2 U}{\partial T^2}\right) = 4\pi G\left(\sigma_0 + \frac{\sigma_1}{c^2}\right) + O(\varepsilon^4). \tag{6.25}$$

Clearly, this unique equation for the approximate gravitational fields of the two first orders, $U$ and $A$, is not enough – the more so, as the same occurs for the dynamical equations, of course. In order to get a solvable approximation scheme at the $1/c^2$ level, one must postulate at least one separate equation including $1/c^2$ contributions, *e.g.*

$$\Delta U = -4\pi G\rho + O(\varepsilon^4). \tag{6.26}$$

Fock [1, §68 *sqq.*] uses implicitly the equivalent of this minimum postulate for the PNA of GR. Other authors postulate two separate equations [2-5], which is actually equivalent. However, Eq.(6.26) is not a consequence of Eq.(6.25). In fact, Eq.(6.26) is not generally true if $\rho$ is assumed to admit a usual expansion $\rho = \rho_0 + \rho_1/c^2 + O(\varepsilon^4)$: (6.26) leads then, unless $\rho_1 = 0$, to the impossible equality

$$\Delta U + 4\pi G\rho_0 = \mathrm{ord}(\varepsilon^2) + O(\varepsilon^4). \tag{6.27}$$

Now if the independent fields $f$, $p$, $\mathbf{u}$ admit asymptotic expansions, Eqs.(6.1)-(6.3), then $\rho$ does admit an expansion, given by Eq.(6.13bis). Further, at the initial time $T = 0$, we may easily calculate its coefficients (6.13bis) as functions of the initial data (5.15). In the varying units used to derive the expansions, $p$ changes to $p' = \varepsilon^{-4}p$, so that (5.15)$_1$ becomes

$$p'_\varepsilon(\mathbf{x},0) = p_{\mathrm{data}}(\mathbf{x}) \equiv \varepsilon_0^{-4} p(\mathbf{x},0) \qquad (\forall \varepsilon), \tag{6.28}$$

where the notations of Section 5 have been adopted, *i.e.*, $p(\mathbf{x},0)$ is the pressure field in the physically given gravitating system (that corresponds to the value $\varepsilon_0$ for $\varepsilon$) at time $T = 0$, in the starting (fixed) units. Let us now come back to the notations of Section 6, thus again omitting the prime and the index $\varepsilon$ for $p$ in the l.h.s., and using the index for ordering the coefficients of the expansions. It follows from (6.28) that the asymptotic expansion of the pressure $p$ is simply, at the initial time:

$$p(\mathbf{x},0) = p_0(\mathbf{x},0) + p_1(\mathbf{x},0)/c^2 + O(\varepsilon^4), \qquad p_0(\mathbf{x},0) = p_{\text{data}}(\mathbf{x}), \ p_1(\mathbf{x},0) = 0. \qquad (6.29)$$

Using Eqs.(6.5) and (6.13bis), we then get

$$\rho_0(\mathbf{x},0) = F_1(p_{\text{data}}(\mathbf{x})), \ \rho_1(\mathbf{x},0) = [\rho_0(\mathbf{u}_0^2/2 + U)](\mathbf{x},0). \qquad (6.30)$$

Equation (6.30)$_2$ definitely proves that $\rho_1 \neq 0$ at the initial time, hence also in some interval $[0,T_0]$ for $T$, and so Eq.(6.26) is definitely incorrect in the frame of the usual method of asymptotic expansions. From (6.25), one may only derive

$$\Delta U = -4\pi G \rho + O(\varepsilon^2), \qquad (6.31)$$

instead of (6.26). Although Eq.(6.31) [together with Eq.(6.25) and the corresponding approximate equations of motion] is sufficient to calculate the post-Newtonian corrections, it is Eq.(6.26) that is required in order to sum the Newtonian contribution and those corrections so as to get a total post-Newtonian acceleration, correct up to and including the $O(\varepsilon^2)$ terms. In other words: if one contents oneself with the asymptotically correct Eqs. (6.25) and (6.31), it does not help to add the post-Newtonian corrections to the "Newtonian" (zeroth-order) contribution, for the latter involves an error which is of the same magnitude as those corrections. According to the method of asymptotic expansions, the equations of the zeroth order, Eqs.(6.9), (6.15) and (6.20), are exact equations (as well as the equations of any order), but they govern the zero-order expanded fields, not the exact fields.

We conclude that *the standard PNA does not pertain to the usual method of asymptotic expansions.* The latter method leads instead to Eqs.(6.9), (6.10), (6.15), (6.18), (6.20) and (6.22), which differ from the equations derivable from the standard PNA in a crucial point: in the former system of equations, one has *two* independent fields $\rho_0$ and $\rho_1$ to characterize the rest-mass density in the second approximation, as well as two fields $\mathbf{u}_0$ and $\mathbf{u}_1$ for the velocity. *We do not mean that the standard PNA is incorrect:* it might find a justification in a different framework from the present framework, which is the usual method of asymptotic expansions applied to a one-parameter family of gravitating systems built from the physically given system. Similarly, the fact that this same framework does not allow to develop an approximation scheme involving wave equations for the gravitational potentials and valid in the near zone, does not mean that the standard PMA is incorrect when one applies it in the near zone. Finally, we emphasize that the present results are still far from allowing rigorous numerical estimates of the error done when one substitutes the expanded fields and equations to the exact ones. At the present stage, one just may expect that the error done with the first PNA, as envisaged here, is equal to a "moderate number" $\lambda$ [$0.01 \leq \lambda \leq 100$ ?] times the next power of the small parameter, thus $\lambda \varepsilon^4$ – for the value $\varepsilon_0$ of the small parameter that corresponds to the physically given gravitating system, which is such that $\varepsilon_0^2 \approx 10^{-6}$ as far as the solar system is concerned.

## 7. Boundary conditions for the expanded fields

The boundary conditions for the expanded fields should be deduced from, or at least compatible with, the initial conditions (5.14)-(5.15) that define the family $(S_\varepsilon)$ of systems. In §6.2, we already found initial conditions for the expansions of the fields $p$ and $\rho$, and these are Eqs. (6.29)$_{2\text{-}3}$ and (6.30)$_{1\text{-}2}$. Similarly, rewriting (5.15)$_2$ in the varying units, we find an initial condition for the expansion of the velocity field **u** :

$$\mathbf{u}_0(\mathbf{x}, T'=0) = \mathbf{u}_{\text{data}}(\mathbf{x}) \equiv \varepsilon_0^{-1} \mathbf{u}_{\varepsilon_0}(\mathbf{x}, T=0), \qquad \mathbf{u}_1(\mathbf{x}, T'=0) = \mathbf{0}, \tag{7.1}$$

where $\mathbf{u}_{\varepsilon_0}$ is the velocity field in the physically given system $S_{\varepsilon_0}$ (expressed in the fixed units), whereas $\mathbf{u}_0$ and $\mathbf{u}_1$ are the coefficients in the expansion (6.2).

Now let us determine the boundary conditions for the gravitational potentials $U$ and $A$. We adopt the explicit notation of Section 5 and begin by rewriting (6.1) in that notation:

$$f'_\varepsilon(\mathbf{x},T') = 1 - 2U(\mathbf{x},T')/c'^2 - 2A(\mathbf{x},T')/c'^4 + O(\varepsilon^6), \tag{7.2}$$

whence (under the relevant uniformity assumption for the expansion):

$$\partial_{T'} f'_\varepsilon(\mathbf{x},T') = -2\partial_{T'} U(\mathbf{x},T')/c'^2 - 2\partial_{T'} A(\mathbf{x},T')/c'^4 + O(\varepsilon^6). \tag{7.3}$$

On the other hand, rewriting (5.14) in the varying units, we get (recall that $\xi = \varepsilon/\varepsilon_0$ and $\varepsilon = c/c'$):

$$f'_\varepsilon(\mathbf{x},0) = 1 - \frac{2}{c'^2}\frac{V(\mathbf{x},0)}{\varepsilon_0^2}, \qquad \frac{\partial f'_\varepsilon}{\partial T'}(\mathbf{x},0) = -\frac{2}{c'^2}\frac{1}{\varepsilon_0^3}\frac{\partial V}{\partial T}(\mathbf{x},0). \tag{7.4}$$

Identifying the powers in $1/c'$, we obtain the initial conditions:

$$U(\mathbf{x}, T'=0) = V(\mathbf{x},T=0)/\varepsilon_0^2, \qquad \partial_T U(\mathbf{x}, T'=0) = \partial_T V(\mathbf{x},T=0)/\varepsilon_0^3, \tag{7.5}$$

$$A(\mathbf{x}, T'=0) = 0, \qquad \partial_T A(\mathbf{x}, T'=0) = 0. \tag{7.6}$$

(By the way, we note that, *had we chosen an exponent n ≠ 3 for $\xi$ in (5.14)$_2$, the orders in (7.3) and in (7.4)$_2$ would not have matched.*) However, the equations for $U$ and $A$, Eqs. (6.9) and (6.10) [with $\sigma_0$ and $\sigma_1$ given by (6.14)], are Poisson equations, hence elliptic, thus the natural boundary conditions for $U$ and $A$ are spatial-boundary conditions at infinity. No such condition is provided by (7.5) and (7.6), except at time $T'=0$. We must hence provide independently the spatial-boundary conditions at infinity for $U$ and $A$. Since we want to describe a spatially-compact gravitating system, we expect that the usual conditions for the Newtonian potential are relevant and thus postulate:

$$(\forall T') \quad U(\mathbf{x}, T') = O(1/r) \text{ and } |\nabla U(\mathbf{x}, T')| = O(1/r^2) \quad \text{as } r \equiv |\mathbf{x}| \to \infty, \tag{7.7}$$

$$(\forall T') \quad A(\mathbf{x}, T') = O(1/r) \text{ and } |\nabla A(\mathbf{x}, T')| = O(1/r^2) \quad \text{as } r \equiv |\mathbf{x}| \to \infty. \tag{7.8}$$

Since the expansion (7.2) is in particular valid for the physical system of interest, characterized by $\varepsilon = \varepsilon_0$, Eqs. (7.7) and (7.8) represent a *constraint* imposed on that system – more precisely, a constraint imposed on the field $f_{\varepsilon_0}(\mathbf{x}, T)$. In fact this seems to be the *second* constraint we have to impose on system $S_{\varepsilon_0}$ {the first one is the requirement that $\varepsilon_0$ [Eq.(5.13)] be finite and indeed small}: even the spatial compactness is not necessary to state the initial-value problem (5.14)-(5.15) that defines the one-parameter family of fields. And the existence of the expansions (6.1)-(6.3), which has been postulated in this work, is likely to be a mathematical consequence of the definition (5.14)-(5.15). Thus, the imposition of the boundary conditions (7.7) and (7.8) has an important physical meaning, for it is precisely at this point that we choose to describe a "quasi-Newtonian" situation.

## 8. Conclusions

**i)** The standard post-Newtonian approximations (PNA) cannot be regarded as asymptotic expansions in the usual sense (§6.2). The same is true for the standard post-Minkowskian approximations (PMA): the method of asymptotic expansions leads to a scheme differing from the standard PNA and PMA schemes in that, according to the former method, all fields (including the velocity and the mass density) must be expanded. The resulting ("asymptotic") PNA scheme is thus, in a sense, more complex than the standard PNA and PMA schemes, but it has an obvious status from the viewpoint of the approximation theory (although mathematical work would be needed, *e.g.,* to prove that the obtained PN expansions are indeed asymptotic to the investigated families of solutions of the exact equations − as seems likely). Furthermore, it leads to local equations that are *linear* with respect to the fields of the approximation following the first ("Newtonian") approximation.

**ii)** In the scalar theory at least (but in fact also in GR), due to the fact that the relevant asymptotic approximation is a post-Newtonian one with instantaneous propagation, the propagation of gravity at speed $c$ has no effect, not only at the order $1/c$, but even up to and including at least the order $1/c^3$ [since we have obtained expansions with a remainder term $O(1/c^4)$]. This is in contrast to what happens in models that *a priori* assume some "aberration formula" for the gravitational force [31-32]. That the model used by Van Flandern [32] leads to enormous effects of propagation, is due to its first-order departure from NG. It turns out that things go differently in relativistic theories of gravitation. The cancellation of first-order retardation effects in a such theory was first obtained by Poincaré [33] (see also Logunov [34]): he actually imposed this cancellation condition, as a constraint on his Lorentz-invariant modification of Newton's attraction law.

**iii)** In the investigated scalar theory, there is actually no non-Newtonian effect at all below the order $1/c^2$. Expansions in even powers of $1/c$ have been postulated here, because in appropriate units the actual small parameter that enters the equations is $1/c^2$, and they have been found entirely consistent. However, we emphasize that, if one postulates instead first-order expansions in $1/c$, one still finds that NG is recovered up to the order $1/c^2$ [18].

**iv)** Since several mass densities intervene in the "asymptotic" schemes proposed, one may expect that the (monopole) mass of a celestial body (*a*) will not be characterized, in those schemes, by only one number, but by several – *e.g.* by two numbers $M_0^a$ and $M_1^a$, at the first PNA. This may be checked by an adequate integration of the present local equations of motion in the volume of the celestial bodies, that provides global equations for the mass centers. This task has now been completed and the results (to be presented in a forthcoming paper) show that the additional masses $M_1^a$ can, however, be expressed in terms of the Newtonian fields of the time $T = 0$ (*i.e.,* the initial time in the initial-value problem considered in Section 5). Now the masses are much like adjustable parameters in celestial mechanics, and one must generally expect that the adjustment is theory-dependent (simply because the equations of motion for the mass centers are theory-dependent). This means that the Newtonian masses $M_N^a$, which are obtained by an adjustment based on the first (zero-order) approximation *alone,* are indeed correct only for Newtonian celestial mechanics: they coincide with the "true" masses of the first approximation, $M_0^a$, only up to an error which is *a priori* of the same order as the second-approximation corrections [21]. Therefore, it would be incorrect to *a priori* dismiss the investigated scalar theory on the ground that it predicts preferred-frame effects in celestial mechanics. But it remains to see if this theory actually provides a more accurate celestial mechanics than does the Newtonian theory.

## 9. Bibliographical note

Preliminary results of this work have been presented in Ref. 18. The first version of the present paper was submitted in February 1999 to another Journal, and rejected. It was pointed out to the author that the construction of a family of similar gravitating systems, like $(S_\varepsilon)$, had been previously studied, within GR, by Futamase & Schutz [35]. (This was not known to the author.) They also use the "weak-gravitational-field limit in Newtonian gravity", called by them the "Newtonian scaling theorem", and they also expand the matter fields as well as the gravitational field. With this in mind, the main new results of the present paper are the following ones:

**i)** The scalar theory is considered, instead of GR.

**ii)** In Section 5 here, the construction of a family of gravitating systems is deduced from the data of a general gravitating system with compact support, assumed to obey the gravitational equations. In contrast, Eqs.(3.13)$_{4-5}$ of Ref. 35 mean that the spatial metric is extremely particular at time $T = 0$, and does not seem appropriate to describe a general gravitating system. Moreover, we have shown that the demand of a Newtonian limit, as defined by conditions (i) and (ii) in §5.3, leads in a rather compelling way to our definition (5.14)-(5.15) of the family of systems. We have discussed the relation between the initial-value problem that defines the family, and the boundary conditions imposed on the expanded fields.

**iii)** We have introduced variable units (depending on the small parameter $\varepsilon$), in which the derivation of asymptotic expansions is straightforward and elementary.

**iv)** Due to the simpler structure of the scalar theory, it has been possible to precisely compare the "standard" PNA scheme with the "asymptotic" PNA scheme and to conclude that the standard PNA cannot be regarded as an asymptotic expansion in the usual sense. Although Futamase & Schutz [35], as well as Rendall [8], emphasized that their ("asymptotic") PNA equations differed from those of the standard PNA in that they expand also the matter fields (in contrast to the standard PNA where only the gravitational field is expanded), they did not examine in detail the consequences as to the compatibility between the "asymptotic" and standard PNA of GR.